\documentclass[onecolumn,showpacs,preprintnumbers,amsmath,amssymb,5pt]{revtex4}

\usepackage{graphicx}% Include figure files
\usepackage{dcolumn}% Align table columns on decimal point
\usepackage{bm}% bold math
\usepackage{multirow}
\usepackage{indentfirst}
\begin{document}

\title{Boltzmann-Equation Based Derivation of Balance Laws in Irreversible Thermodynamics}% Force line breaks with \\
\author{Liu Hong*, Zaibao Yang, Yi Zhu, Wen-An Yong}
\affiliation{Zhou Pei-Yuan Center for Applied Mathematics, Tsinghua University,
Beijing, China, 100084}
\email{zcamhl@tsinghua.edu.cn}

\begin{abstract}
In this paper we propose a novel approach to construct macroscopic balance equations and constitutive equations describing various irreversible phenomena. It is based on the general principles of non-equilibrium thermodynamics and consists of four basic steps: picking suitable state variables, choosing a strictly concave entropy function, separating entropy fluxes and production rates properly, and determining a dissipation matrix. Our approach takes the advantage of both EIT and GENERIC formalisms, and shows a direct correspondence with Levermore's moment closure hierarchies for the Boltzmann equation. This result may put various macroscopic modeling approaches starting from the general principles of non-equilibrium thermodynamics on a solid microscopic foundation based on the Boltzmann equation.
\end{abstract}

\maketitle

\section{Introduction}
\setlength{\parindent}{3em}
To derive thermodynamically admissible constitutive relations is a central task of non-equilibrium thermodynamics. In the past decades, various macroscopic thermodynamic approaches, \textit{e.g.} Classical Irreversible Thermodynamics (CIT), Rational Thermodynamics (RT), Extended Irreversible Thermodynamics (EIT) and GENERIC formalism, have achieved tremendous successes in the mathematical modeling of natural phenomena \cite{Groot, Jou, Ottinger}, such as polymeric fluids, heat transfer, nano materials, ultrasonic waves, second sounds, \textit{etc}.. A common feature of those methods is that they are all starting from the general principles of macroscopic thermodynamics. No details in the microscopic dynamics have been considered during the modeling.

Alternatively, there is another group of so-called ``kinetic approaches'', which seek to construct models based on the microscopic kinetics. Their staring points usually are the master equation, the Langevin equation, the Fokker-Planck equation, the Boltzmann equation and so on. Then with some moment-closure methods, time-evolution equations for the macroscopic moments can be directly derived from the microscopic kinetics. A most remarkable example dues to the Chapman-Enskog expansion in the 1910s\cite{Kremer}, based on which the classical Navier-Stokes equations, a most famous governing model for the macroscopic fluid mechanics, were derived from the Boltzmann equation for the first time. Later on, a lot of different approaches based on the Boltzmann equation were proposed for the same purpose, \textit{e.g.} the maximal entropy principle\cite{Kogan}, the Grad's\cite{Grad} and the Levermore's hierarchies\cite{Levermore}. Now the kinetic approach has already become a standard way to link the macroscopic dynamics with the microscopic kinetics. However, there is no guarantee that the derived macroscopic equations in such a way will be thermodynamically admissible.

In this paper, we propose a novel approach to construct macroscopic balance equations and constitutive equations based on the general principles of non-equilibrium thermodynamics. The four basic elements of our approach: picking suitable state variables, choosing a strictly concave entropy function, separating entropy fluxes and production rates properly, as well as determining a dissipation matrix, are illustrated step by step. Most importantly from our illustration, a direct correspondence between our macroscopic thermodynamical approach and the Levermore's moment closure hierarchies for the Boltzmann equation could be clearly seen. This fact actually points out a possibility to place various macroscopic modeling approaches starting from the general principles of non-equilibrium thermodynamics on a solid microscopic foundation based on the Boltzmann equation -- a long-unsolved problem in this field. Although most previous macroscopic thermodynamic approaches claim their roots in the kinetic theory, this point has never been clearly justified before.

\section{Four Elements of Our Approach}
Inspired by EIT and GENERIC formalisms, our approach also starts with the general principles of non-equilibrium thermodynamics: the conservation of the total energy and the positiveness of entropy production rates. There are four basic elements in our procedure: (i) choosing suitable state variables (including both conserved and dissipative quantities) which can fully characterize the system under consideration, (ii) constructing a strictly concave entropy function with respect to the chosen state variables, (iii) selecting the proper entropy flux, as well as (iv) determining a dissipation matrix which characterizes the irreversibility of the system. In the following, we will show how to construct these four elements step by step. In parallel, the Levermore's moment closure hierarchies will be referred to from time to time. And its underlying correlations with our macroscopic thermodynamic approach will be revealed and discussed in each step.

For the convenience of following discussions, we want to present a very brief introduction to the Levermore's moment closure hierarchies for the Boltzmann equation before proceeding to our main results. It is well-known that the Levermore's method provides a nice coarse-graining way to obtain the desired macroscopic dynamic equations from the microscopic kinetics characterized by the Boltzmann equation. It reads as follows \cite{Levermore,Muller}
\begin{eqnarray}
\frac{\partial}{\partial t}\int c_ifd\vec{\xi}+\nabla\cdot\int\vec{\xi}c_ifd\vec{\xi}=\int c_i(f_{\star}f'_{\star}-ff')B(\vec{\omega},\vec{\xi},\vec{\xi}')d\vec{\omega} d\vec{\xi}'d\vec{\xi},\qquad i=0,1,\cdots,n-1.
\end{eqnarray}
Here $\int c_ifd\vec{\xi}$ is referred to as the $i$-th order moment and $\{c_i=c_i(\vec{\xi}), i=0,1,\cdots, n-1\}$ constitutes an admissible space in the sense of \cite{Levermore}. In (1), $\vec{\xi}$ and $\vec{\xi}'$ are the velocities of particles before binary collisions, $\vec{\xi}_{\star}$ and $\vec{\xi}'_{\star}$ are the velocities of particles after collision, $f=f(\vec{r},\vec{\xi},t)$, $f'=f'(\vec{r},\vec{\xi}',t)$, $f_{\star}=f_{\star}(\vec{r},\vec{\xi}_{\star},t)$ and $f'_{\star}=f'_{\star}(\vec{r},\vec{\xi}'_{\star},t)$ denote the respective distribution functions of particles in the phase space, $\vec{\omega}$ denotes the scattering angle, and $B(\vec{\omega},\vec{\xi},\vec{\xi}')=B(\vec{\omega},\vec{\xi}',\vec{\xi})=B(\omega,\vec{\xi}_{\star},\vec{\xi}'_{\star})$ is the collision kernel and is positive almost everywhere in its domain \cite{Cercignani}. Let $\alpha_i=\alpha_i(\vec{r},t)$ ($i=0,1,\cdots,n-1$) be an $i$-th order unknown tensor and substitute the Ansatz
\begin{eqnarray}
f(\vec{r},\vec{\xi},t)=\exp\big[\sum_{i=0}^{n-1}c_i(\vec{\xi})\odot\alpha_i\big]
\end{eqnarray}
introduced in \cite{Levermore} into (1), we get a set of first-order partial differential equations for the unknown parameters $\alpha_i$, which are Galilean invariant thanks to the admissibility of $\{c_i\}$. Here $\odot$ denotes the tensor product, \textit{i.e.} $A\odot B=\sum_{b_1,b_2,\cdots,b_m}A_{a_1a_2\cdots a_nb_1b_2\cdots b_{m-1}b_m}B_{b_mb_{m-1}\cdots b_2b_1}$. For the Levermore's moment closure hierarchies, the state variables and the entropy function are already given, while the other two steps will be illustrated in what follows.

\subsection{State Variables}
The choice of a suitable set of state variables, which can fully characterized the desired kinetic details of the system under study, is a prerequisite for the mathematical modeling of general non-equilibrium processes.
Using different state variables may lead to different governing
equations, and suitable state variables are expected to give
simple governing equations which can directly reveal physical insights
of the processes. Based on their different dynamical behaviors, state variables can be classified into conserved and dissipative ones. Conserved variables are relatively simple and determined through the underlying conservation laws or symmetry properties of the system, while the choice of dissipative variables is usually none unique and much harder. Till now, there is no general rule for this purpose. EIT suggested to use the unknown variables appeared in the conservation equations directly\cite{Jou}. But in our recent works, we found that the usage of conjugated variables with respect to the pre-specified entropy may be better in some way\cite{Zhu}.

Motivated by the moment-closure method mentioned above, here we consider an isolated thermodynamic system and assume that it can be fully characterized by state variables $\phi\equiv\{\phi_0,\phi_1\cdots,\phi_{n-1}\}$. The subscripts $0$ to $n-1$ denote the different tensor orders of the state variables, like the scalar, vector, tensor and so on. However, how to properly choose these state variables is not a major task of our current study. Interested readers may refer to the related works we mentioned above.

These variables correspond to the moments defined in (1) as $\phi_i=\int c_ifd\vec{\xi}$ ($i=0,1,\cdots,n-1$) with $\{c_i\}$ an admissible space in the sense of \cite{Levermore}, ensuring the Galilean invariance of the derived moment-closure system. Actually, through these normalization conditions, the unknown parameters $\alpha_i$ in the Ansatz (2) can be well determined in terms of the $\phi_i$'s. In the kinetic theory, the moments up to order three are usually taken as the density $\rho=\int fd\vec{\xi}$, the momentum $M_i=\int \xi_ifd\vec{\xi}$, the second-order stress tensor $P_{ij}=\int \xi_i\xi_jfd\vec{\xi}$ and the third-order heat flux tensor $Q_{ijk}=\int \xi_i\xi_j\xi_kfd\vec{\xi}$, where the particle mass is assumed to be one.

\subsection{Entropy function}
In the next step, we need to pick an entropy function $S=S(\phi_0,\phi_1,\cdots,\phi_{n-1})$, which is strictly concave with respect to the pre-chosen state variables $\phi_i$ ($i=0,1,\cdots,n-1$). For this function, there is no further restriction, since it sensitively depends on the details of the problem to be modeled. In the literature, the Boltzmann entropy and the Tsallis entropy are two most widely referred forms. The former adopts a log form and is addable, while the latter takes a polynomial form and is non-addable\cite{Tsallis}. As a special example, we may take $S=-k_B\int (f\ln f-f)d\vec{\xi}$ in accordance with the Boltzmann equation, where $f$ is defined in (2).

Set $S_{\phi_i}\equiv(\partial S/\partial\phi_i)_{\phi_0,\cdots,\phi_{i-1},\phi_{i+1},\cdots,\phi_{n-1}}$. We have the generalized Gibbs relation \cite{Jou}
\begin{eqnarray}
dS=\sum_{i=0}^{n-1}S_{\phi_i}^T\odot d\phi_i
\end{eqnarray}
and therefore the evolution equation for entropy
\begin{eqnarray}
\frac{\partial S}{\partial t}=\sum_{i=0}^{n-1}S_{\phi_i}^T\odot \frac{\partial\phi_i}{\partial t}=-\nabla\cdot \vec{J}^S+\sigma^S
\end{eqnarray}
with
\begin{eqnarray}
\sigma^S\equiv\nabla\cdot\vec{J}^S+\sum_{i=0}^{n-1}S_{\phi_i}^T\odot \frac{\partial\phi_i}{\partial t}.\nonumber
\end{eqnarray}
Here $\vec{J}^S$ stands for the entropy flux and $\sigma^S$ is the entropy production rate, which should always be non-negative according to the second-law of thermodynamics\cite{Groot, Jou}.

A proper choice of the entropy flux $\vec{J}^S$ is crucial for the modeling. In some previous studies, the entropy flux has not been paid enough attention to and is taken for granted\cite{Jou}. Here we treat this part more carefully. We split the entropy flux into two parts, \textit{i.e.} $\vec{J}^S=\vec{J}^S_1+\vec{J}^S_2$. Inspired by the moment method for the Boltzmann equation, we take the first part as
\begin{eqnarray}
\vec{J}^S_1&=&\sum_{i=0}^{n-1}\phi_{i+1}\odot S_{\phi_i}^T,
\end{eqnarray}
where the unknown tensor $\phi_n$ now can be regarded as the coefficient of $S_{\phi_{n-1}}$ in the entropy flux $\vec{J}^S_1$. Later, it will be seen that $\phi_n$ is the flux of the state variable $\phi_{n-1}$. The second part of the entropy flux $\vec{J}^S_2$ will be determined in the next section. Now the entropy production rate becomes
\begin{eqnarray}
\sigma^S=\sum_{i=0}^{n-1}S_{\phi_i}^T\odot\bigg(\frac{\partial\phi_i}{\partial t}+\nabla\cdot\phi_{i+1}\bigg)+\sum_{i=0}^{n-1}\nabla S_{\phi_i}^T\odot\phi_{i+1}+\nabla\cdot\vec{J}^S_2.
\end{eqnarray}

\subsection{Entropy flux}
Since the first part is fixed, the entropy flux will be totally determined if we can properly set its second part. A direct way, which we will see is also quite meaningful both mathematically and physically, is to require the sum of last two terms in above formula (6) to be vanished, \textit{i.e.}
\begin{eqnarray}
\sum_{i=0}^{n-1}\nabla S_{\phi_i}^T\odot\phi_{i+1}+\nabla\cdot\vec{J}^S_2=0.
\end{eqnarray}
Or equivalently,
\begin{eqnarray}
\sum_{i=0}^{n-1}S_{\phi_i\phi_j}^T\odot\phi_{i+1}+\partial\vec{J}^S_2/\partial\phi_j=0,\qquad \forall j=0,\cdots,n-1,
\end{eqnarray}
dues to the generality of state variable $\phi_j$. It is a central equation to determine the two unknowns $\phi_n$ and $\vec{J}^S_2$ as functions of $(\phi_0,\phi_1,\cdots,\phi_{n-1})$. In such a way, the n-th order tensor $\phi_n$ can be expressed in terms of other known lower-order state variables and obtain the so-called constitutive equations. Till now, we only know that above equation admits at least on solution (see the discussions below on the Boltzmann equation). However, whether this solution is unique or not is an open problem. If (7) admits other solutions, one would have alternative ways for moment closure and thus other possibilities of constitutive equations. Further discussions are beyond the scope of this paper.

It is easily seen that the requirement imposed by Eq. 7 is equivalent to the well-known entropy condition for conservation laws $\partial\phi_i/\partial t+\nabla\cdot\phi_{i+1}=0$ by multiplying $\nabla S_{\phi_i}^T$ on both sides. This entropy condition dues to Godunov, Friedrichs, Lax and \textit{et al.} \cite{Gudunov,Friedrichs,Boillat} and was firstly proposed for symmetrical hyperbolic systems of first-order PDEs. It is well recognized in mathematics that hyperbolicity is a substantial requirement for systems of first-order PDEs to be well-posed\cite{Serre}. Based on our current macroscopic thermodynamic approach, the hyperbolicity of the resultant PDEs is clearly verified with the help of (7), while no other extended theory (including EIT and GENERIC) till now can reach this point in general.

In the case of the Boltzmann equation, as $f=\exp(\sum_{i=0}^{n-1}c_i\odot\alpha_i)$, $\phi_i=\int c_ifd\vec{\xi}$ ($i=0,1,\cdots,n-1$) and $S=-k_B\int (f\ln f-f)d\vec{\xi}$, it is easy to verify that
\begin{eqnarray}
&&\sum_{i=0}^{n-1}\nabla S_{\phi_i}^T\odot\phi_{i+1}=\sum_{i=0}^{n-1}\nabla \bigg(\sum_{j=0}^{n-1}\frac{\partial S}{\partial \alpha_j}\frac{\partial \alpha_j}{\partial \phi_i}\bigg)^T\odot\phi_{i+1}\nonumber\\
&=&\sum_{i=0}^{n-1}\nabla \bigg[-k_B\sum_{j=0}^{n-1}\bigg(\sum_{k=0}^{n-1}\alpha_k\int c_jc_kfd\vec{\xi}\bigg)\bigg(\int c_ic_jfd\vec{\xi}\bigg)^{-1}\bigg]^T\odot\bigg(\int \vec{\xi}c_ifd\vec{\xi}\bigg)\nonumber\\
&=&\sum_{i=0}^{n-1}-k_B\nabla\alpha_i^T\odot \frac{\partial \vec{B}}{\partial\alpha_i}=-k_B\nabla\cdot\vec{B},\nonumber
\end{eqnarray}
where $\vec{B}=\int \vec{\xi}fd\vec{\xi}$. Therefore the entropy condition (7) is solved by $\phi_n=\int \vec{\xi}c_{n-1}fd\vec{\xi}$ and $\vec{J}^S_2=k_B\int \vec{\xi}fd\vec{\xi}$.

\subsection{Dissipation Matrix}
To keep the entropy production rate be non-negative ($\sigma^s\geq0$) in consistence with the second-law of thermodynamics, it is quite natural to take
\begin{eqnarray}
\frac{\partial\phi_i}{\partial t}+\nabla\cdot\phi_{i+1}=\sum_{j=0}^{n-1}\lambda_{\phi_i\phi_j}S_{\phi_j},\qquad i=0,\cdots,n-1
\end{eqnarray}
in which the dissipation matrix $\bm{\lambda}=(\lambda_{\phi_i\phi_j})$ is symmetric and positive semi-definite. Furthermore, we require the null space of $\bm{\lambda}$ to be independent of the $\phi_i$'s. As pointed out in \cite{Yong}, this requirement is important for the stability of non-equilibrium processes. Formulas in (9) are known as the balance equations. Together with the constitutive equations, they constitute a closed system once the entropy function $S$ and the dissipation matrix $\bm{\lambda}$ are specified.

Now we show that the collision integrals on the right-hand side of Levermore's moment closure hierarchies for the Boltzmann equation can be reformulated into the form of dissipative entropic flows defined in (9). To do this, we substitute the Ansatz (2) into (1) and define
\begin{eqnarray}
Q(\vec{\alpha})\equiv\int \vec{c}^T(\vec{\xi})\big\{\exp[\vec{c}(\vec{\xi}_{\star})\cdot\vec{\alpha}+\vec{c}(\vec{\xi}'_{\star})\cdot\vec{\alpha}]-\exp[\vec{c}(\vec{\xi})\cdot\vec{\alpha}+\vec{c}(\vec{\xi}')\cdot\vec{\alpha}]\big\}Bd\vec{\omega} d\vec{\xi}'d\vec{\xi},
\end{eqnarray}
where $\vec{c}(\vec{\xi})=(c_0,c_1,\cdots,c_{n-1})$, $\vec{\alpha}=(\alpha_0,\alpha_1,\cdots,\alpha_{n-1})^T$ and $\vec{c}(\vec{\xi})\cdot\vec{\alpha}=\sum_{k=0}^{n-1}c_k(\vec{\xi})\odot\alpha_k(\vec{r},t)$. With the celebrated identity \cite{Cercignani}
\begin{eqnarray}
4\int c_i(\vec{\xi})(f_{\star}f'_{\star}-ff')Bd\vec{\omega} d\vec{\xi}'d\vec{\xi}=\int\big[c_i(\vec{\xi})+c_i(\vec{\xi}')-c_i(\vec{\xi}_{\star})-c_i(\vec{\xi}'_{\star})\big](f_{\star}f'_{\star}-ff')Bd\vec{\omega} d\vec{\xi}'d\vec{\xi},
\end{eqnarray}
which holds for any continuous function $c_i(\vec{\xi})$, we can rewrite
\begin{eqnarray}
Q(\vec{\alpha})=-\frac{1}{4}\int[\vec{c}^T(\vec{\xi}_{\star})+\vec{c}^T(\vec{\xi}'_{\star})-\vec{c}^T(\vec{\xi})-\vec{c}^T(\vec{\xi}')\big]\big\{\exp[\vec{c}(\vec{\xi}_{\star})\cdot\vec{\alpha}+\vec{c}(\vec{\xi}'_{\star})\cdot\vec{\alpha}]-\exp[\vec{c}(\vec{\xi})\cdot\vec{\alpha}+\vec{c}(\vec{\xi}')\cdot\vec{\alpha}]\big\}Bd\vec{\omega} d\vec{\xi}'d\vec{\xi}.
\end{eqnarray}
It follows from the mean-value theorem that
\begin{eqnarray}
\exp[\vec{c}(\vec{\xi}_{\star})\cdot\vec{\alpha}+\vec{c}(\vec{\xi}'_{\star})\cdot\vec{\alpha}]-\exp[\vec{c}(\vec{\xi})\cdot\vec{\alpha}+\vec{c}(\vec{\xi}')\cdot\vec{\alpha}]=M[\vec{c}(\vec{\xi}_{\star})+\vec{c}(\vec{\xi}'_{\star})-\vec{c}(\vec{\xi})-\vec{c}(\vec{\xi}')]\cdot\vec{\alpha},
\end{eqnarray}
where
\begin{eqnarray}
M(\vec{\alpha},\vec{\xi},\vec{\xi}',\vec{\xi}_{\star},\vec{\xi}'_{\star})=\int_0^1\exp\big\{\tau[\vec{c}(\vec{\xi}_{\star})+\vec{c}(\vec{\xi}'_{\star})-\vec{c}(\vec{\xi})-\vec{c}(\vec{\xi}')]\cdot\vec{\alpha}+[\vec{c}(\vec{\xi})+\vec{c}(\vec{\xi}')]\cdot\vec{\alpha}\big\}d\tau.
\end{eqnarray}
Then the dissipation matrix is found to be
\begin{eqnarray}
\bm{\lambda}(\vec{\alpha})=\frac{1}{4}\int[\vec{c}^T(\vec{\xi}_{\star})+\vec{c}^T(\vec{\xi}'_{\star})-\vec{c}^T(\vec{\xi})-\vec{c}^T(\vec{\xi}')\big]MB\big[\vec{c}(\vec{\xi}_{\star})+\vec{c}(\vec{\xi}'_{\star})-\vec{c}(\vec{\xi})-\vec{c}(\vec{\xi}')\big]d\vec{\omega} d\vec{\xi}'d\vec{\xi},
\end{eqnarray}
which is symmetric and positive semi-definite. Moreover, it is well-known that mass, moment and energy are the only three conservation laws for the Boltzmann equation \cite{Grad}, which follows that the null space $\{\vec{\alpha}\in R^n:\int|[\vec{c}(\vec{\xi}_{\star})+\vec{c}(\vec{\xi}'_{\star})-\vec{c}(\vec{\xi})-\vec{c}(\vec{\xi}')]\cdot\vec{\alpha}|^2d\vec{\omega}d\vec{\xi}'d\vec{\xi}=0\}$ of $\bm{\lambda}(\vec{\alpha})$ is independent of $\phi$. Thus the collision integrals are
\begin{eqnarray}
Q(\vec{\alpha})=-\bm{\lambda}(\vec{\alpha})\cdot\vec{\alpha}=\bm{\lambda}(\vec{\alpha}(\vec{\phi}))\cdot S_{\vec{\phi}},
\end{eqnarray}
by noticing the identity $\vec{\alpha}=-S_{\vec{\phi}}$.

\section{Orthogonal Conditions}
Now we try to show that the general principles of non-equilibrium thermodynamics are fulfilled in our formulism. Notice that above balance equations could be rewritten into a compact form
\begin{eqnarray}
\frac{\partial\vec{\phi}}{\partial t}=\vec{J}^\phi_c+\vec{J}^\phi_d.
\end{eqnarray}
Here the vector for state variables is
\begin{eqnarray}
\vec{\phi}=(\phi_0,\phi_1,\cdots,\phi_{n-1})^T,
\end{eqnarray}
the conserved energetic flows are
\begin{eqnarray}
\vec{J}^\phi_c&=&-\nabla\cdot\big(\phi_{1},\phi_{2},\cdots,\phi_{n}\big)^T,
\end{eqnarray}
and the dissipative entropic flows are
\begin{eqnarray}
\vec{J}^\phi_d&=&\bm{\lambda}\odot S_{\vec{\phi}}.
\end{eqnarray}

In view of the kinetic theory, we write the density $\rho=\phi_0$, the momentum $\vec{M}=\rho \vec{v}=\phi_1$, the second-order stress tensor $\textbf{P}=\phi_2$ and the third-order heat flux tensor $\textbf{Q}=\phi_3$, and introduce $U=tr\textbf{P}/2-\vec{M}^2/(2\rho)$ as the internal energy. Then following relations can be verified, \textit{i.e.}
\begin{eqnarray}
U_{\vec{\phi}}^T\cdot \vec{J}^\phi_c&=&-\nabla\cdot(U\vec{v})-\nabla\cdot\vec{q}-\textbf{p}^T:\nabla\vec{v},\\
U_{\vec{\phi}}^T\cdot \vec{J}^\phi_d&=&\frac{1}{2}v_i^2\sum_{j=0}^{n-1}\lambda_{\rho\phi_j}\odot S_{\phi_j}-v_i\sum_{j=0}^{n-1}\lambda_{M_i\phi_j}\odot S_{\phi_j}+\frac{\delta_{ki}}{2}\sum_{j=0}^{n-1}\lambda_{P_{ik}\phi_j}\odot S_{\phi_j},
\end{eqnarray}
where $p_{ij}=P_{ij}-\rho v_iv_j$ and $q_i=Q_{jji}/2-P_{ij}v_j-P_{jj}v_i/2+\rho v_j^2v_i$ are recognized as the traditional pressure and heat flux tensors, respectively. To recover the first law of thermodynamics, which states the change in the internal energy is equal to the amount of heat flux plus the amount of work done by its surroundings, terms on the right-hand side of (22) should vanish. Therefore the energetic forces $U_{\vec{\phi}}$, defined as the gradient of the internal energy with respect to state variables, and the dissipative entropic flows are orthogonal. Physically, this orthogonal relation means that the energetic forces make no contribution to the dissipative entropic flows, such that the corresponding state variables will be conserved. Due to the arbitrariness of velocity $v_i$,
\begin{eqnarray}
\sum_{j=0}^{n-1}\lambda_{\rho\phi_j}\odot S_{\phi_j}=\sum_{j=0}^{n-1}\lambda_{M_i\phi_j}\odot S_{\phi_j}=\sum_{j=0}^{n-1}\sum_{i=1}^{3}\lambda_{P_{ii}\phi_j}\odot S_{\phi_j}=0
\end{eqnarray}
are required by the Galilean invariance in general. In this way, the classical conservation laws for mass, momentum and total energy are recovered.

Additionally, we have
\begin{eqnarray}
S_{\vec{\phi}}^T\cdot \vec{J}^\phi_c&=&-\nabla\cdot\vec{J}^S,\\
S_{\vec{\phi}}^T\cdot \vec{J}^\phi_d&=&\sigma^S.
\end{eqnarray}
The first formula shows another orthogonal relation between the conserved energetic flows and the entropic forces (the gradient of entropy with respect to state variables). Physically, it means that the entropic forces do not disturb the conserved energetic flows, which in turn guarantees the conservation of mass, momentum and total energy during the irreversible processes. It is noted that (24) and (25) together guarantee the second law of thermodynamics, namely, the entropy of an isolated system never decreases.

\section{Several Examples}
\subsection{Equilibrium State and the Euler Equations}
As a first example, we look at the equilibrium system characterized by the state variables: the density $\rho$, momentum $\rho\vec{v}$ and total energy $\rho e$, which corresponds to the minimum admissible space \cite{Levermore}. In this case, it is well-known that the solution of the Boltzmann equation is given by the Maxwell-Boltzmann distribution, \textit{i.e}. $f|_e=\rho\big(2\pi k_BT\big)^{-3/2}\exp\big[-(\vec{\xi}-\vec{v})^2/(2k_BT)\big]$, where
$T=(2e-\vec{v}^2)/(3k_B)$ is the temperature. Based on this, we can get the entropy function $S|_e=k_B\rho\big(3\ln T-2\ln\rho+A\big)/2$, the entropic forces $S_{\vec{x}}|_e=\big(S|_e\rho+\vec{v}^2/(2T)-5k_B/2,-v_i/T,\delta_{ij}/(2T)\big)^T$ with $A=3\ln(2\pi k_B)+3$ a constant, the second-order pressure tensor $P_{ij}|_e=\rho v_iv_j+\rho k_BT\delta_{ij}$, and the third-order heat flux tensor $Q_{ijk}|_e=\rho v_iv_jv_k+\rho k_BT(v_i\delta_{jk}+v_j\delta_{ik}+v_k\delta_{ij})$. Accordingly, it is easy to verify that the entropy condition (8) is satisfied, and the entropic fluxes are found to be $\vec{J}^S_1|_e=S|_e\vec{v}-k_B\rho\vec{v}$, $\vec{J}^S_2|_e=k_B\rho\vec{v}$ and $\vec{J}^S|_e=S|_e\vec{v}$. Now the balance equations reduce to the Euler equations
\begin{eqnarray}
&&\frac{\partial}{\partial t}\rho+\nabla\cdot(\rho \vec{v})=0,\nonumber\\
&&\frac{\partial}{\partial t}(\rho\vec{v})+\nabla\cdot(\rho \vec{v}\vec{v})+\nabla p=0,\nonumber\\
&&\frac{\partial}{\partial t}(\rho e)+\nabla\cdot(\vec{v}\rho e+\vec{v}p)=0,\nonumber
\end{eqnarray}
where $p=\rho k_BT$ is known as pressure in equilibrium. Additionally, the evolution equation for the entropy becomes
\begin{eqnarray}
&&\frac{\partial}{\partial t}\big(S|_e\big)=-\nabla\cdot\big(\vec{v}S|_e\big).\nonumber
\end{eqnarray}
This hints that the entropy is conserved in equilibrium.

\subsection{A Gaussian Model with Ten-Moments}
Let's go beyond the equilibrium state and study a Gaussian model with ten-moments. In this case, the state variables are given as the density $\phi_0\equiv\rho$, momentum $\phi_1\equiv\rho\vec{v}$ and symmetric pressure tensor $\phi_2\equiv \textbf{P}$. Specifically, we set the entropy function as $S=k_B\rho[\ln\det\mathbf{\Theta}-2\ln\rho+3\ln(2\pi)+3]/2$, where $\mathbf{\Theta}=\rho^{-1}\textbf{P}-\vec{v}\vec{v}$. It is a direct generalization of the entropy in the equilibrium state. Furthermore, it is noted that the entropy function thus defined is a strictly concave function with respect to $\rho$, $\rho\vec{v}$ and $\textbf{P}$.

By noticing $S_{\phi_0\phi_0}/k_B=-\rho^{-1}\vec{v}\vec{v}:\mathbf{\Theta}^{-1}-\rho^{-1}-\rho/2[\mathbf{\Theta}^{-1}:(\partial\mathbf{\Theta}/\partial{\rho})]^{2}$, $S_{\phi_0\phi_1}/k_B=\rho^{-1}\vec{v}\cdot\mathbf{\Theta}^{-1}+\vec{v}\cdot\mathbf{\Theta}^{-2}:(\partial\mathbf{\Theta}/\partial{\rho})$, $S_{\phi_0\phi_2}/k_B=-1/2\mathbf{\Theta}^{-2}:(\partial\mathbf{\Theta}/\partial{\rho})$, $S_{\phi_1\phi_1}/k_B=-\rho^{-1}\mathbf{\Theta}^{-1}-2\rho^{-1}\vec{v}\vec{v}:\mathbf{\Theta}^{-2}$, $S_{\phi_1\phi_2}/k_B=\rho^{-1}\vec{v}\cdot\mathbf{\Theta}^{-2}$ and $S_{\phi_2\phi_2}/k_B=-1/2\rho^{-1}\mathbf{\Theta}^{-2}$, we see the entropy condition (8) turns to be
\begin{eqnarray}
&&k_B\{-\rho^{-1}\vec{v}\vec{v}:\mathbf{\Theta}^{-1}-\rho^{-1}-\rho/2[\mathbf{\Theta}^{-1}:(\partial\mathbf{\Theta}/\partial{\rho})]^{2}\}\rho\vec{v}+k_B[\rho^{-1}\vec{v}\cdot\mathbf{\Theta}^{-1}+\vec{v}\cdot\mathbf{\Theta}^{-2}:(\partial\mathbf{\Theta}/\partial{\rho})]\cdot\textbf{P}\nonumber\\
&&+k_B[-1/2\mathbf{\Theta}^{-2}:\partial\mathbf{\Theta}/\partial{\rho}]:\phi_3+\partial\vec{J}^S_2/\partial \rho=0,\nonumber\\
&&k_B[\rho^{-1}\vec{v}\cdot\mathbf{\Theta}^{-1}+\vec{v}\cdot\mathbf{\Theta}^{-2}:(\partial\mathbf{\Theta}/\partial{\rho})]\rho\vec{v}+k_B[-\rho^{-1}\mathbf{\Theta}^{-1}-2\rho^{-1}\vec{v}\vec{v}:\mathbf{\Theta}^{-2}]\cdot\textbf{P}\nonumber\\
&&+k_B[\mathbf{\Theta}^{-2}:\partial\mathbf{\Theta}/\partial{(\rho\vec{v})}]:\phi_3+\partial\vec{J}^S_2/\partial (\rho\vec{v})=0,\nonumber\\
&&k_B[-1/2\mathbf{\Theta}^{-2}:(\partial\mathbf{\Theta}/\partial{\rho})]\rho\vec{v}+k_B[\rho^{-1}\vec{v}\cdot\mathbf{\Theta}^{-2}]\cdot\textbf{P}+k_B[-1/2\mathbf{\Theta}^{-2}:\partial\mathbf{\Theta}/\partial{\textbf{P}}]:\phi_3+\partial\vec{J}^S_2/\partial \textbf{P}=0.\nonumber
\end{eqnarray}
It can be verified that above equations are solved by $(\phi_3)_{ijk}=v_iP_{jk}+v_{j}P_{ik}+v_kP_{ij}-2\rho v_iv_jv_k$ and $\vec{J}^S_2=k_B\rho\vec{v}$ (in fact, the solution to above equations permits an additional term $\mathbf{\Theta}^2:\partial \vec{g}(\mathbf{\Theta})/\partial\mathbf{\Theta}$ in $\phi_3$ and corresponding $\vec{g}(\mathbf{\Theta})/2$ in $\vec{J}^S_2$, where $\vec{g}(\mathbf{\Theta})$ is an arbitrary vector function of $\mathbf{\Theta}$). Now the balance equations become
\begin{eqnarray}
&&\frac{\partial}{\partial t}\rho+\nabla\cdot(\rho \vec{v})=0,\nonumber\\
&&\frac{\partial}{\partial t}(\rho\vec{v})+\nabla\cdot \textbf{P}=0,\nonumber\\
&&\frac{\partial}{\partial t}\textbf{P}+\nabla\cdot\phi_3=\frac{k_B}{2}\lambda_{\textbf{P}\textbf{P}}:\mathbf{\Theta}^{-1}.\nonumber
\end{eqnarray}
Additionally, we can determine the entropic fluxes as $\vec{J}^S_1=S\vec{v}-k_B\rho\vec{v}$ and $\vec{J}^S=S\vec{v}$. Therefore the evolution equation for the entropy is given by
\begin{eqnarray}
&&\frac{\partial}{\partial t}S=-\nabla\cdot\big(\vec{v}S\big)+\frac{k_B^2}{4}\mathbf{\Theta}^{-1}:\lambda_{\textbf{P}\textbf{P}}:\mathbf{\Theta}^{-1}.\nonumber
\end{eqnarray}

Now we are going to show how above balance equations reduce to the Euler equations in the equilibrium state. We split the pressure tensor into two parts $\textbf{P}=2\textbf{E}+\mathbf{\tau}$, where $\textbf{E}=(p\textbf{I}+\rho\vec{v}\vec{v})/2$, then the last two balance equations become
\begin{eqnarray}
&&\frac{\partial}{\partial t}(\rho\vec{v})+\nabla\cdot \textbf{P}=\frac{\partial}{\partial t}(\rho\vec{v})+\nabla\cdot(\rho\vec{v}\vec{v})+\nabla p+\nabla\cdot{\mathbf{\tau}}=0,\nonumber\\
&&\frac{\partial}{\partial t}\textbf{P}+\nabla\cdot\phi_3=2\bigg[\frac{\partial}{\partial t}\textbf{E}+\nabla\cdot\bigg(\vec{v}\textbf{E}+\frac{1}{3}\vec{v}p\textbf{I}\bigg)\bigg]+\bigg[\frac{\partial}{\partial t}\mathbf{\tau}+\nabla\cdot\big(\vec{v}\mathbf{\tau}\big)+(\nabla\vec{v})\cdot\mathbf{\tau}+\mathbf{\tau}\cdot(\nabla\vec{v})^T\nonumber\\
&&\qquad\qquad\qquad\quad+\vec{v}\nabla\cdot\mathbf{\tau}+(\nabla\cdot\mathbf{\tau})\vec{v}\bigg]=\rho k_B\lambda_{\textbf{P}\textbf{P}}:\big(p\textbf{I}+\mathbf{\tau}\big)^{-1}.\nonumber
\end{eqnarray}
Take trace on both sides of the last equation and notice the fact that $trE|_e=(3p/2+\rho v^2/2)|_e=\rho e$, $\mathbf{\tau}|_e=0$ and $\lambda_{P_{ii}P_{jj}}=0$, we will arrive at the Euler equations in the equilibrium state. Furthermore, it is easy to verify that the entropy function defined above for the ten-moments reduces to the classical one $S|_e$ in the equilibrium state.

Interestingly, above macroscopic formulation has a perfect microscopic correspondence. If we take the distribution function as the multi-dimensional gaussian distribution, \textit{i.e.} $f=\rho(2\pi)^{-3/2}(\det\mathbf{\Theta})^{-1/2}\exp\big[-(\vec{\xi}-\vec{v})\cdot\mathbf{\Theta}^{-1}\cdot(\vec{\xi}-\vec{v})/2\big]$, then it can be shown that the entropy function defined above is given by the Boltzmann formula $S=-k_B\int (f\ln f-f)d\vec{\xi}$. While the macroscopic state variables are in fully consistent with the Levermore's Ansatz, \textit{i.e.} $\rho=\int fd\vec{\xi}$, $\rho\vec{v}=\int \vec{\xi}fd\vec{\xi}$, $\textbf{P}=\int \vec{\xi}\vec{\xi}fd\vec{\xi}$ and $\phi_3=\int\vec{\xi}\vec{\xi}\vec{\xi}fd\vec{\xi}$.

\section{Conclusion}
As a summary, in this paper we have proposed a novel thermodynamic approach to construct the macroscopic balance equations and constitutive equations. Our approach takes the advantage of both EIT and GENERIC formalisms, and shows a complete consistency with the general principles of non-equilibrium thermodynamics. More importantly, the underlying connections between our macroscopic derivation and Levermore's moment closure hierarchies for the Boltzmann equation are fully explored. This interesting result may put various macroscopic modeling approaches, starting from the general principles of non-equilibrium thermodynamics, on a solid microscopic foundation based on the Boltzmann equation. A long-standing problem has been tacitly assumed to be correct but not clearly justified.

Towards Levermore's moment closure hierarchies for the Boltzmann equation, four basic steps in our derivation: the choice of suitable state variables, the construction of a strictly concave entropy function, the proper separation of entropy flux and entropy production rate, as well as the determination of the dissipative structure have been illustrated step by step. Due to the close linkage between microscopic kinetics and macroscopic dynamics, our derivation may shed light on other macroscopic modeling approaches for irreversible thermodynamics, in which a major bottleneck is how to properly construct those four elements.\\

\section*{Acknowledgment}
This work was supported by the Tsinghua University Initiative Scientific Research Program (Grants 20121087902
and 20131089184) and by the National Natural Science Foundation of China (Grants 11204150 and 11471185).

\begin{figure}[h]
\includegraphics[width=9.0cm]{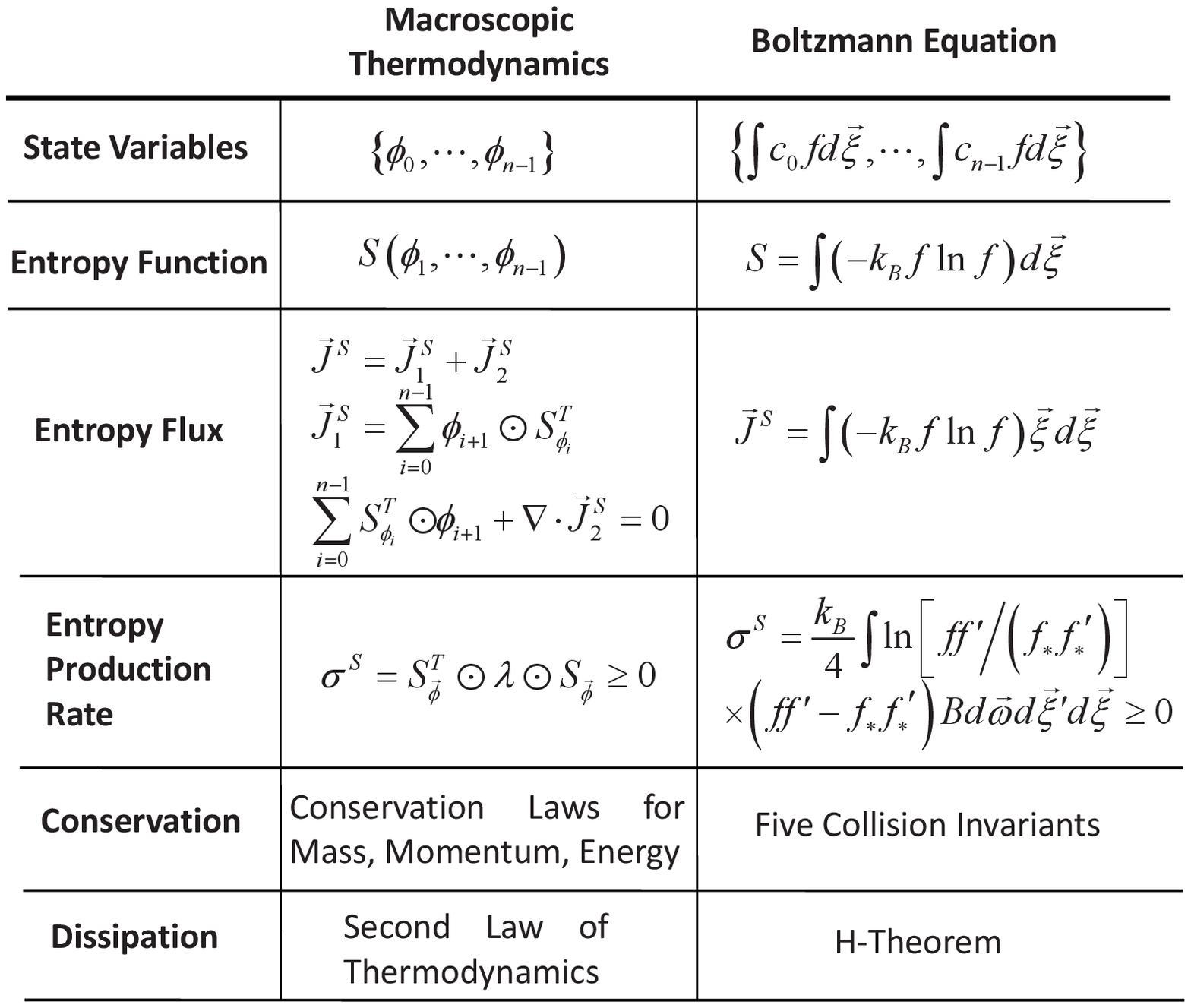}% Here is how to import EPS art
\caption{The correspondence between our macroscopic thermodynamic approach and the moment closure hierarchies for the Boltzmann equation.}
\end{figure}

\end{document}